**Title**

# Growth and Structure of alpha-Ta films for Quantum Circuit Integration

**Authors**

Loren D. Alegria[1]*, Alex Abelson[1], Eunjeong Kim[1], Soohyun Im[2], Paul M. Voyles[3], Vincenzo Lordi[1], Jonathan L Dubois[1], Yaniv J. Rosen[1]

**Affiliations**

[1] Physics Division, Lawrence Livermore National Laboratory, Livermore, CA, 94550, USA

[2] Department of Materials Design and Innovation, University at Buffalo, Buffalo, NY, 14260, USA

[3] Department of Materials Science and Engineering, University of Wisconsin-Madison, Madison, WI, 53706, USA

*corresponding author: alegria4@llnl.gov

**Abstract**

Tantalum films incorporated into superconducting circuits have exhibited low surface losses, resulting in long-lived qubit states. Remaining loss pathways originate in microscopic defects which manifest as two level systems (TLS) at low temperature. These defects limit performance, so careful attention to tantalum film structures is critical for optimal use in quantum devices. In this work, we investigate the growth of tantalum using magnetron sputtering on sapphire, Si, and photoresist substrates. In the case of sapphire, we present procedures for growth of fully-oriented films with $\alpha$-Ta [1 1 1] // $Al_2O_3$ [0 0 0 1] and $\alpha$-Ta [1 -1 0] // $Al_2O_3$ [1 0 -1 0] orientational relationships, and having residual resistivity ratios (RRR) ~ 60 for 220 nm thick films. On Si, we find a complex grain texturing with Ta [1 1 0] normal to the substrate and RRR ~ 30. We further demonstrate airbridge fabrication using Nb to nucleate $\alpha$-Ta on photoresist surfaces. For the films on sapphire, resonators show TLS-limited quality factors of $1.3 \pm 0.3 \times 10^6$ at 10 mK (for waveguide gap and conductor widths of 3 $\mu$m and 6 $\mu$m, respectively). Structural characterization using scanning electron microscopy, X-ray diffraction, low temperature transport, secondary ion mass spectrometry, and transmission electron microscopy reveal the dependence of residual impurities and screw dislocation density on processing conditions. The results provide practical insights for fabrication of advanced superconducting devices including qubit arrays, and guide future work on crystallographically deterministic qubit fabrication.

**Text**

### *Introduction*

Studies of two level systems (TLS) in superconducting qubit devices have sought the microscopic origins of qubit decoherence.[1-8] Although atomic displacements in junction oxides are the best understood TLS, recent studies highlighting the role of surface oxides, hydrides and microfractures have illustrated the complexity of the problem.[9-11] Overall, most posited decoherence sources argue for the development of superconductor films with high quality oxides and uniform crystal structure.

Recent demonstrations of high-performance qubits fabricated with tantalum as the wiring layer have drawn interest to materials science considerations particular to this superconductor.[12,13] The apparent performance advantage over more conventional metals (such as Al and Nb) has been attributed to both the surface cleanliness enabled by tantalum's resistance to corrosive cleaning agents (e.g., piranha solution) and also its favorable native oxide, which may be more nearly stoichiometric and therefore contain fewer magnetic impurities than Nb surface oxides.[12-15] The growth of superconducting Ta films with well-defined microstructure is therefore of paramount interest for the fabrication of qubits with minimal dissipation. Nonetheless, it is not obvious that simply more crystalline films will be preferable for quantum circuits made from Ta. For example, less crystalline films may possess superior vortex pinning and reduced substrate intermixing, while more crystalline films may contain less oxide.

Substrate requirements for superconducting circuits include minimal piezoelectricity, low impurity concentrations, and suitable processing stability. Samples of heat-exchange-growth sapphire and float-zone silicon, as used here, have low-temperature loss tangents suitable for millisecond-scale qubit lifetimes.[16-19]

Tantalum has two distinct polymorphs. The body-centered-cubic (bcc) $\alpha$-Ta has a bulk superconducting critical temperature of $T_C$ = 4.4 K. On the other hand, $\beta$-Ta, a tetragonal σ-phase with $T_C$ ~ 0.5 K, is typically disordered but forms preferentially on many substrates at room temperature. Experiments to date have indicated that the $\beta$ phase results in greater losses in the few-photon limit relevant to superconducting qubits, motivating our investigation of ordered $\alpha$-Ta films.[20]

Growth of Ta on sapphire structurally follows the model of Nb.[20-24] Sapphire permits epitaxial growth at the elevated temperatures at which the $\alpha$-phase forms readily.[25] Like Nb, coherent epitaxy of bcc Ta can occur on the a- and c-planes of sapphire.[22,26] For 100 nm-scale film thicknesses as often used for circuits, the [110] Ta films that form on the sapphire a-plane show uniaxially elongated grain texture.[27,28] On the c-plane, Ta also can grow in the [110] direction, but exhibits significant lattice mismatch, leading to the formation of columnar grains which are elongated along the sapphire [1 0 -1 0], [0 1 -1 0], or [1 1 -2 0] directions.[12,29] At higher temperatures or reduced background pressure, Ta grows in the [111] direction with a single azimuthal orientation, $\alpha$-Ta [1 -1 0] // Al$_2$O$_3$ [1 0 -1 0], equivalent to $\alpha$-Ta (-1 1 0) // Al$_2$O$_3$ (2 -1 -1 0).[30] The lattice mismatch between the trigonal interfacial lattices is 1.9 % as illustrated in Fig. 1A.[30-32]

On the other hand, studies of growth on silicon surfaces have not seen crystallographic registry but $\alpha$-Ta grows at temperatures above approximately 300 °C as well as at very low temperatures (~ 7 K).[29,33,34] At low temperatures, the Si-Ta interfacial phases are suppressed but the density of grain boundaries is high. Silicon's advantages in manufacturability may outweigh any negative impact of these dislocations on qubit performance.[33]

Finally, growth of superconducting Ta on organic surfaces is of interest for airbridge definition or for the creation of other free-standing structures.[35-38] Airbridges allow superconducting circuits to overcome the severe topology constraints of planar circuits. Typically made from Ti or Al, airbridges are widely used to accurately define resonant modes and waveguides. Although possibly less constrained than qubit materials, airbridges should have low loss,[36] and ideally withstand the same surface treatments, such as piranha cleaning, as the base superconductor layer.

The three-dimensional structure of the airbridge can be defined using a range of template materials, but photoresist templates afford the most flexibility and are particularly suitable for Ta, since it is compatible with piranha etchant, which can remove organic residues that have been cross-linked by the sputtering process.  Below we demonstrate that a 2 nm layer of Nb suffices to promote phase-pure α-Ta on a photoresist substrate, permitting straightforward fabrication of superconducting Ta airbridges.

*Growth Studies*

We studied the growth of Ta on Si, sapphire, and photoresist surfaces. With the exception of sample Sph 0, which was grown by a commercial vendor, all the films were grown by confocal sputtering with continuous 20 rpm substrate rotation under 0.4 Pa Argon at a throw distance of 150 mm from 3 inch sources from a base pressure of $5\times10^{-8}$ Torr.  The growth rate was proportional to DC sputtering power, and was 0.2 nm/s for 270 W.

Table 1 summarizes results of Ta depositions on sapphire using a range of magnetron sputtering deposition parameters.   We examined the parameters' effect on the film resistivity, which is a useful proxy for the density of scattering sites in the film, including the presence of impurity β-Ta secondary phase.  Here, resistance was measured by taking several (usually three) four-terminal probe measurements at a fixed distance from the center of the 4" wafer until three significant figures were established.

| ID | $T_A$ [°C] | $T_G$ [°C] | Nucl. | $P_{DC}$ [W] | $\rho$ [μΩ-cm] | $R_{300K}/R_{5K}$ | XRD | $\delta_{110}$ (deg.) |
|---|---|---|---|---|---|---|---|---|
| Sph 0* | - | 500 | no | - | 23 | 7.8 | α-110 | 0.21 |
| Sph 1 | 25 | 25 | yes | 270 | 27.3 | 2.6 | α-110 | 0.28 |
| Sph 2 | 25 | 25 | yes | 270 | 27.7 | - | - | - |
| Sph 3 | 650 | 625 | yes | 150 | 16.2 | - | - | - |
| Sph 4 | 825 | 800 | no | 150 | 16.9 | - | α-111 | $\delta_{111}$ = 0.24 |
| Sph 5 | 825 | 800 | no | 150 | 16.1 | - | - | - |
| Sph 6 | 650 | 625 | no | 270 | 18.3 | - | - | - |
| Sph 7 | 650 | 625 | no | 270 | 16.1 | - | - | - |
| Sph 8 | 650 | 625 | no | 150 | 15.9 | - | - | - |
| Sph 9 | 650 | 625 | no | 150 | 15.2 | - | - | - |
| Sph 10 | 650 | 625 | no | 75 | 14.4 | 62 | α-111 | $\delta_{111}$ = 0.22 |
| Sph 11 | 650 | 625 | no | 40 | 14.5 | - | - | - |
| Sph 12 | 650 | 625 | no | 40 | 14.4 | - | - | - |

**Table 1. Tantalum films on c-plane sapphire.**  For a given wafer, $T_A$ is the *in situ* annealing temperature held for three minutes prior to growth, $T_G$ is the growth temperature, *Nucl*. indicates the use of a 2 nm Nb nucleation layer, $P_{DC}$ is the DC power applied to the 3 inch target, $\rho$ is the room temperature resistivity as measured 20 mm from the center of the 100 mm wafer, XRD is the phase and orientation indicated by XRD, and $\delta_{110}$ is the width of the (110) reflection in XRD.  Room temperature resistivity was strongly indicative of phase and defect density.  Of the parameters varied, the batch index had the clearest impact on resistivity: in sequential, nominally identical growths the second consistently showed lower resistivity (compare Sph 4/5, Sph 6/7, Sph 11/12), indicating that background impurities impact defect density in the films. *Sph 0 was deposited by a commercial vendor and is 200 nm in thickness. All other samples are 220 nm in thickness.

The clearest trend is the effect of the chamber condition.  For example, the sample pairs Sph 4/5, Sph 6/7, Sph 8/9, and Sph 11/12 were each grown identically and immediately in sequence, and, in each case, the second deposit was appreciably more conductive (lower $\rho$). This indicates that with each successive deposition, the chamber environment becomes cleaner, leading to fewer scattering sites in the Ta.  The effect occurred despite conditioning procedures prior to performing growth series.  Such observations are of practical value, as they allow for estimation of the relative effects of growth conditions versus chamber condition on film product.

Following the MBE literature on Ta epitaxy,[25] we anticipated obtaining the highest quality films at the highest growth temperatures, but instead found increased resistivity at 800 °C, possibly due to chamber outgassing.[25,39]  In addition, we expected growth at relatively high rates ~ 0.2 nm/s would reduce background gas effects, but consistently found the lowest resistances during the slowest growths (down to 40 W)  This had diminishing benefits, since using 40 W vs 75 W resulted in minimal resistance change (Sph 10/11/12) and required several hours deposition.  (In a later batch in which the 40 W growth preceded a 75 W growth, the 40 W product was ~ 1 % more conductive.) We note that the film Sph 0, the only film grown off-site, formed in the less conductive [110] orientation, likely due to background gas incorporation in that growth system.[30]  Although the chamber history effect made precise measurement of growth parameter effects challenging, in general we found that the low DC sputtering power, $P_{DC}$ < 150 W and a temperature of 625 °C were close to optimal for highly conductive films, approaching the bulk resistivity (13.5 μΩ-cm).

Next, Table 2 presents illustrative growths performed on Si.  Float-zone Si wafers were dipped in 6:1 buffered oxide etch (BOE) for two minutes prior to deposition.  Samples Si 1 and Si 2 had very high resistivity, indicating β-Ta formation at the low growth temperatures, independent of whether the wafer was annealed *in situ*.  With the inclusion of the nucleation layer (Si 3 and Si 4) the product became much more conductive, consistent with the α-phase as confirmed by XRD of Si 3. Comparing Si 3 and Si 4, nucleated growth at moderately elevated temperatures increases crystallinity as compared to room temperature.  At higher temperatures (Si 5 and 6), a considerably more crystalline film forms with low resistivity.  Comparing Si 5 and Si 6, the film product is entirely independent of Si orientation, with identical resistivity ratios in films on (110)- and (100)-Si (cf. Fig 4B) suggesting that the interfacial layer fully buffers the Si lattice.

| ID | Orient. | $T_A$ [°C] | $T_G$ [°C] | Nucl. | $P_{DC}$ [W] | $\rho$ [$\mu\Omega$-cm] | $R_{300 K}/R_{5 K}$ | XRD | $\delta_{110}$ (deg.) |
|---|---|---|---|---|---|---|---|---|---|
| Si 1 | (100) | 25 | 400 | no | 270 | 290 | - | $\alpha,\beta$ | 0.42 |
| Si 2 | (100) | 650 | 300 | no | 150 | 180 | - | - | - |
| Si 3 | (100) | 25 | 25 | yes | 270 | 34.8 | - | $\alpha$-110 | 0.40 |
| Si 4 | (100) | 650 | 300 | yes | 150 | 19.3 | - | - | - |
| Si 5 | (100) | 650 | 625 | no | 75 | 15.9 | 32 | $\alpha$-110 | 0.23 |
| Si 6 | (110) | 650 | 625 | no | 75 | 15.7 | 32 | $\alpha$-110 | 0.24 |
| Resist | - | 25 | 25 | yes | 270 | 29 | - | $\alpha$-110 | 0.52 |

**Table 2. Tantalum films on silicon and resist substrates**. At lower temperatures (Si 1 - 4) a Nb nucleation layer proved necessary for $\alpha$-phase growth, independent of annealing. At high temperature, the film structure was independent of silicon orientation. Notation as in Table 1.

Comparing the films grown on Si and sapphire, the room temperature resistivity is a reliable indicator of the fractions of $\alpha/\beta$ phase, as others have observed,[23] but only a partial indicator of crystal orientation. Si 5/6 and Sph 4 have nearly identical resistivities, but XRD indicates that Si 5/6 had [110] orientation while Sph 4 had [111]. Ta crystallization is known to be sensitive to background impurities,[30] yet even in the general-purpose sputtering system employed here, proper chamber conditioning and straightforward variation of growth parameters result in films with high conductivity, a critical heuristic for the phase.

***Structural and Chemical Characterization***

Depending on substrate and deposition conditions, magnetron-sputtered $\alpha$-Ta films exhibit a wide variety of surface structures.[40] Fig. 1 serves to illustrate this morphological variety.

Fig. 1A shows the alignment between the $\alpha$-Ta (111) and $Al_2O_3$ (0 0 0 1) surfaces as typically occurs at higher growth temperatures.[30] Such growth temperatures were employed for samples Sph 3-12. Fig. 1B and C show high-angle annular dark-field scanning transmission electron microscopy (HAADF-STEM) images and scanning electron microscopy (SEM) images, respectively, of films grown under these conditions (Sph 10 and Sph 8, respectively). In HAADF-STEM, a very sharp interface between the Ta and sapphire is observed. Grain boundaries were not apparent in the regions of the lamella studied. On the other hand, SEM showed characteristic point-like defects in all such surfaces, as seen in Fig. 1C. The appearance of these is consistent with ubiquitously-observed screw dislocations in bcc metals. Typically possessing ½[111] Burgers vectors, these one dimensional defects appear as points or small triangles where they intersect the surface.[41,42] The alignment of these triangles reflects the azimuthal epitaxial relation. By contrast, Fig. 1D shows an SEM image of a [110]-oriented Ta film (Sph 0) with clear grain boundaries.[29] In this way, the SEM morphology of freshly prepared films can be used to assess the film orientation.

On Si, a more complex grain structure appears. At low growth temperature, grains have random azimuthal orientation,[29] but at elevated growth temperature longer range ordering effects are apparent.  Fig. 1E shows an SEM image of Si 5 and the emergence of alternating bands of wide and narrow grains.  Si 6 showed an effectively identical texture.  The SEM signal contrast here was visible at high (~ 20 kV) beam acceleration, and thus derives from proportionately deep within the film, rather than describing surface topography.  Cross-sectional STEM imaging (Fig S1A) confirmed the surface to be smooth with occasional grain boundaries.  These findings are consistent with previous reports that Ta forms at random in-plane rotations at the scale accessible by XRD,[33] but the aligned bands we observe indicate ordering of the Ta at up to 100 µm scales (c.f. Fig. S1B).  Given the similarity of this length scale to that of qubit lithography, it is possible that the intersection of such bands with qubit electrodes could influence the variability of qubit quality.

After Si and sapphire, polymers represent a third major class of substrate in qubit device fabrication.  But, during sputtering, the impinging, energetic Ta is expected to alter the photoresist surface and, in turn, the Ta nucleation kinetics, so it was far from clear whether $\alpha$-Ta films could be grown on photoresist at all.  In order to study the growth of Ta on photoresist, we deposited Nb-nucleated Ta on both uniform photoresist films and on patterned resist templates for airbridges, with the latter being the most relevant to qubit devices.

Fig. 1F shows an SEM image of the foot of an airbridge grown by nucleation on a photoresist template.  The top of the image is the contact region, which grew on the sapphire, and the bottom of the image is the suspended film, which grew on the photoresist.   In contrast to growth on cleaner sapphire surfaces, the contact crystallites are randomly oriented due to residues from the lithography and oxygen-plasma cleaning steps that create the scaffold.  The contacts show a coarser grain structure than in the bridge span and both show faceting, which is absent in $\beta$-Ta films, indicating the $\alpha$-phase.[29] The phase apparent from SEM was confirmed by XRD and transport as described subsequently.

Fig. 2 shows the procedure for airbridge fabrication.  First, a template structure on sapphire is formed using two AZ1518 photoresist layers.  The first layer of resist is spun to a thickness of 2 µm, baked at 90 °C and patterned into rectangles defining the 'scaffold' of the bridge (Fig. 2B).  The sample is then heated ('reflowed') at 140 °C for five minutes to round the scaffold (Fig. 2C).  Next, another 2 µm layer of AZ1518 is spun and patterned to form the rectangular windows defining the perimeter of the bridge.  The final template structure is shown in the optical image in Fig. 2D.

The Ta deposition on this photoresist template is identical to the process described in Ref. [29]:  2 nm of Nb is sputtered at ambient temperature, followed by 200 nm of Ta.  The wafer is then soaked in acetone, removing the metal that was not within the airbridge perimeter (liftoff).  After rinsing in isopropanol and drying with nitrogen, the airbridges appear as in Fig. 2E-F. Here, Fig. 2E is an SEM image captured at an angle and shows that the scaffold was fully removed. (The bridges shown in Fig. 2E are the ones created by the masks in the dashed area of Fig. 2D.)  Fig. 2F shows an optical microscope image exhibiting a stubborn residue around the perimeter of the scaffold.  These particular bridges shared a common scaffold, so the residue forms lines joining the four bridges.  This wafer was immersed in a fuming piranha bath (3 parts of concentrated sulfuric acid and 1 part 30 wt. % hydrogen peroxide) for 1 hour.   It was then rinsed in deionized water and ultrasonically cleaned in isopropanol for 1 minute.  After this cleaning step the organic residues were completely removed as seen in Fig. 2F.

We found that it was possible to directly characterize the phase of suspended bridges by room-temperature transport. To do so, we performed four-terminal resistance measurements of devices similar to that of Fig. 2H, with and without the presence of the bridge. We thereby obtained precise measurements of the resistance of a given bridge, and combined this with SEM-determined dimensions of the curved structure to calculate the resistivity. The bridges we analyzed had a resistivity of 29 +/- 1 $\mu\Omega$-cm. Purely $\beta$-phase films would have a resistivity approximately an order of magnitude higher, allowing us to conclude that the bridge is $\alpha$-phase with a high degree of confidence. Films with similar resistivity (e.g. Si 3) have $T_C$ = 4.1 K. In addition, XRD of films grown under the same conditions but on a uniform, 2 $\mu$m thick, 90 °C-baked AZ1518 photoresist layer show no detectable $\beta$ fraction, as evident in Fig. 3. These two indicators show that bridges formed in this way are alpha phase.

In Fig. 3 we compare the $\theta$-$2\theta$ XRD spectra of a representative subset of the films grown on Si, Sapphire, and photoresist. The [110]-oriented films on sapphire (Sph 0, 1) show a set of peaks between 30 and 35 which we ascribe to strain around grain boundaries.[12,29] On the other hand, $\alpha$-Ta films on silicon (Si 1,3,5,6) show no such extra peaks, only the (110) reflection ($2\theta$ = 38.5°). Growth at low temperature (below 500° C, Sample Si 1) shows a clear $\beta$-(002) peak as well ($2\theta$ = 33.7°). The Ta film nucleated on photoresist, labeled '*Resist,*' shows no $\beta$ or $\alpha$-(111) contribution. Finally, in contrast to all the latter samples, the only Ta-related reflection in the [111]-oriented Sph 10 and 4 films is the (222) plane ($2\theta$ = 107.8°) with no $\alpha$-(110), minor peaks, or $\beta$ peaks. The complete XRD spectra, including the $\alpha$-(220) and $\alpha$-(002) reflections are included in Fig. S2, and further illustrates the high degree of crystallinity in the [111]-oriented films.

The quality of the films grown on sapphire at 625 °C temperature was borne out in variable temperature resistance measurements. The film Sph 10 showed $T_C$ = 4.3 K, close to the bulk value of $T_C$ = 4.48 K, and residual resistivity ratio, RRR = 62, almost double that of the films grown on Si under the same conditions (c.f. Ref. [43]), and almost 10 times that of [110] films on sapphire. The RRR provides a sensitive measure of the defect density.

The lower $T_C$ and RRR of the films on Si (Fig. 4) may be compared to the STEM data of Fig. S3. Here an 8-10 nm boundary layer is apparent between the Si and Ta. STEM-based energy dispersive spectroscopy (EDS) indicated the boundary layer to smoothly interpolate between pure Si and Ta (to within the ~ 1 %), and scattering within the disordered region might explain the lower RRR. Nonetheless, the RRR in the films grown on Si were much higher than in the fine-grained, nucleated films (such as Sph 3 in Fig. S3A).

Although STEM shows the Ta films grown at high and low temperatures to have clear crystalline phases, secondary ion mass spectrometry (SIMS) reveals a substantial quantity of contaminants in the films, as shown in Fig. 5. In these data the light elements have an overall calibrated concentration accuracy of ~ 15 %, with sample-to-sample accuracy of ~ 5 %. For the metals, the Ta bulk is taken to be 100% Ta and the sapphire to be 40 % Al.

An important consideration for the SIMS data is that the films grown at room temperature had a substantially rougher surface (~10 nm in roughness) than the films grown at 625 °C. The measured concentration profile represents a convolution of this surface roughness with the true local concentration. Even in light of this, H, C, and O are present much further beneath the surface in the room-temperature-grown film than the higher temperature-grown film (c.f. Figs. 6A and 6B). Percolation of light elements along grain boundaries and the volatility of many C, H, and O-bearing species at the higher temperature explain this data.  In the case of Sph 10 (Fig. 5C), the $10^{19}$ cm$^{-3}$ scale of carbon impurities in the film interior is not far from the nominal impurity concentration of the 99.9 % Ta sputtering target.

Finally, the high conductivity film, Sph 10, was patterned into hanger-mode, frequency-multiplexed coplanar waveguide λ/4 resonators, using the mask design of Ref. [44] The design features coupling quality factors $Q_C$ = 5 x10$^5$ and metal-substrate, substrate-air, metal-air, and substrate participation ratios of 0.26,  0.13, 0.008, and 89 respectively.  The resonators were defined with a reactive ion etch process (ULVAC ICP 500 W RF, 100 W DC, 1 Pa, 40 sccm SF$_6$, 60 s) and then cleaned with a fresh piranha solution (20 minutes) and 6:1 buffered oxide etch (5 minutes).  After the etch (which strips the native oxide) they were exposed to atmospheric conditions for approximately four hours before being characterized in a dilution refrigerator at 10 mK using a vector network analyzer.

Fig. 6 summarizes the results of the resonator measurements.  Two resonators on the chip had very low quality factors, and were not analyzed.  The resonance line shapes of each resonator were measured as a function of power and were fit to the asymmetrical Lorentzian model[45] to calculate $Q_C$ and $Q_i$, the coupling and internal quality factor, at each power.  While $Q_C$ is determined by the resonator-feedline coupling geometry, $Q_i$ reflects the losses within the resonator at that power.  The $Q_C$ obtained were independent of power and consistent with the design value.  The $Q_i$ (and the deviations, $\sigma_{Qi}$) obtained from the regression are plotted in Fig. 6 as a function of photon number, $n = 2PQ_L^2/(\hbar Q_c (2\pi f_0)^2)$, where $P$ is the incident power at the sample, $Q_L$ is the resonator loaded quality factor, and $f_0$ is the resonator frequency.  The $Q_i(n)$ data is then fit according to the form $Q_i^{-1}= Q_{TLS}^{-1}(1+n/n_C)^{-\beta/2} + Q_h^{-1}$ where $Q_{TLS}$ is the TLS-limited internal quality factor, $n_C$ is the critical photon number, $\beta$ is a phenomenological exponent, and $Q_h$ is the quality factor in the high power limit.[46]  The fits give a mean $Q_{TLS}$ = 1.3 ± 0.3 x10$^6$, independent of fitting details.  (The other fit parameters are included in the Supplemental Material.)  The wide range of $Q_{TLS}$ illustrates the susceptibility of such measurements to inhomogeneities introduced during microfabrication, but the overall low loss indicates that such films are suitable as the primary superconducting layer for present high-performance qubit circuits.[16]

**Conclusion**

We have demonstrated suitable growth conditions for high-quality Ta films with low densities of scattering defects on technologically important substrates, including Si, sapphire, and photoresist. The results are applicable to many of the diverse microfabrication applications of Ta films, including qubit fabrication.[47]  For qubits, the demonstration of growth on polymer substrates allows for fabrication of piranha-compatible airbridges, enabling a convenient workflow for multi-qubit fabrication.[48]

Further work is required to clearly evaluate any merits of such airbridges as compared to more conventional Al bridges, as by comparative studies of resonators where Al or Ta bridges span the resonators or are part of the center conductor. The latter case is of growing importance given the widespread use of flyover structures in present qubit circuits which remain an uncertain source of loss.[49]  Additionally, the nature of the bridge underside is of continued interest, recommending experimentation with alternative nucleating species.[50]

As regards growth on Si, Ta/Si interfacial mixing uniformly depresses the stability of the superconducting state for Ta films grown on Si, encouraging the exploration of low-loss diffusion barriers, and innovative approaches to low temperature growth.[51]  Studies of silicide-containing resonators formed deliberately at higher temperature are also needed to guide qubit design.[52]

In general, correlating any new materials concept with qubit circuit performance will be facilitated by co-fabrication of diverse test device types alongside qubits, each type aimed at isolating a given loss mechanism (e.g. bridge under-side losses). This is because many studies to date have examined the impact of a given material change upon overall qubit performance with little observed statistical significance, because the qubit itself did not happen to be limited by the loss mechanism of interest.  Automated analysis of such built-in metrology devices will allow for clear separation of loss channels, even for the most unconventional qubit geometries or materials.

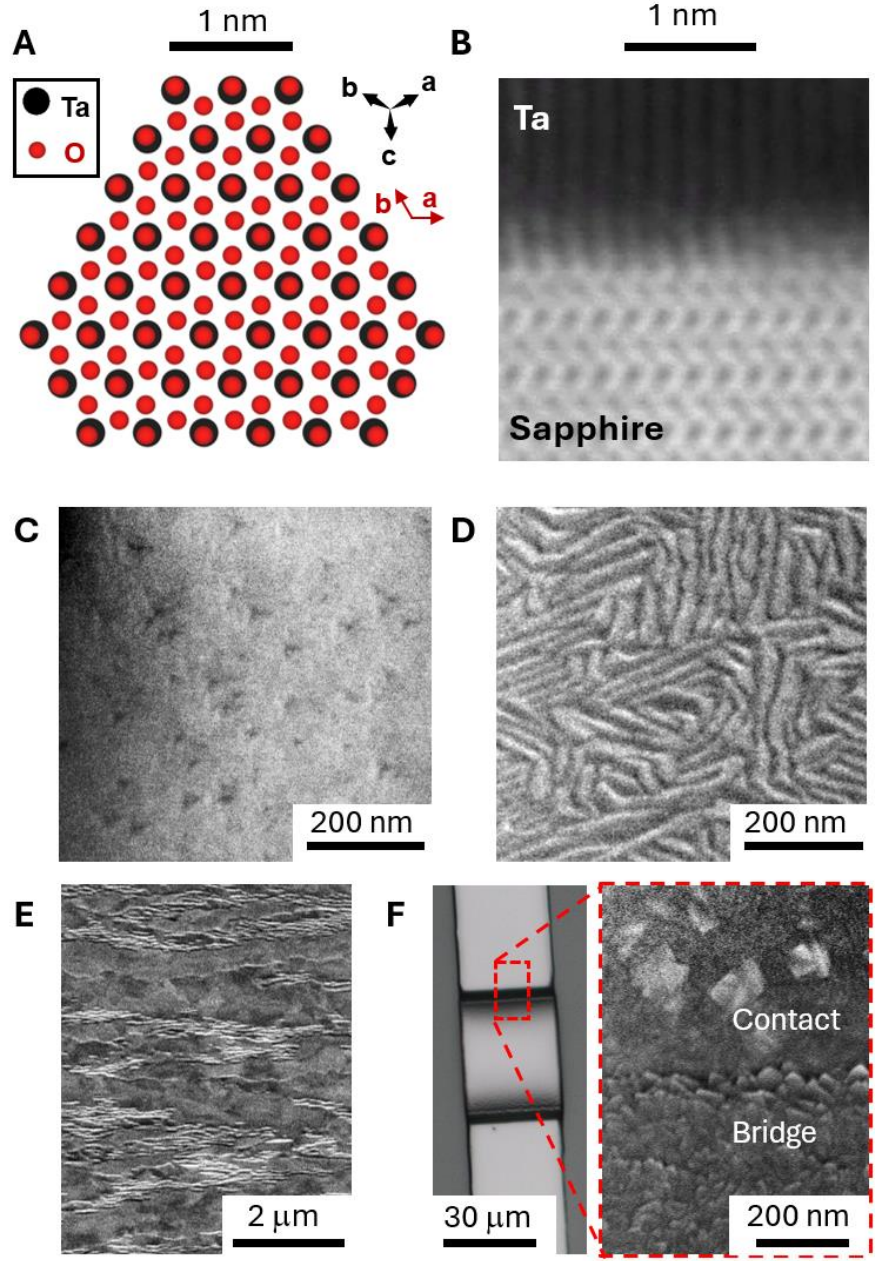

**Figure 1. Characteristic microstructures of α-Ta grown on sapphire, Si, and photoresist.** (A) The (001) sapphire oxygen bulk loci approximately match the (111) Ta surface. (B) HAADF-STEM image of a cross section of a [111]-Ta film on sapphire, showing an abrupt interface between the substrate and metal. (C) SEM image of a [111]-oriented Ta film surface, grown on a sapphire substrate. Screw dislocations are visible as dark, triangular spots. (D) By contrast, SEM of a [110]-Ta film on sapphire shows distinct grain boundaries. (E) On silicon, Ta films grown at 625 °C show a banded grain texture, independent of the Si lattice. (F) Optical (*left*) and SEM (*right*) images of a Nb-nucleated Ta airbridge, in which the growth substrate is photoresist. The SEM image of the contact-bridge interface region shows crystallites indicative of α-Ta in both the contact and bridge. [Images B-E are of samples Sph 10, Sph 8, Sph 0, Si 5; note that images C-E are of the film surface at 60° tilt to increase contrast].

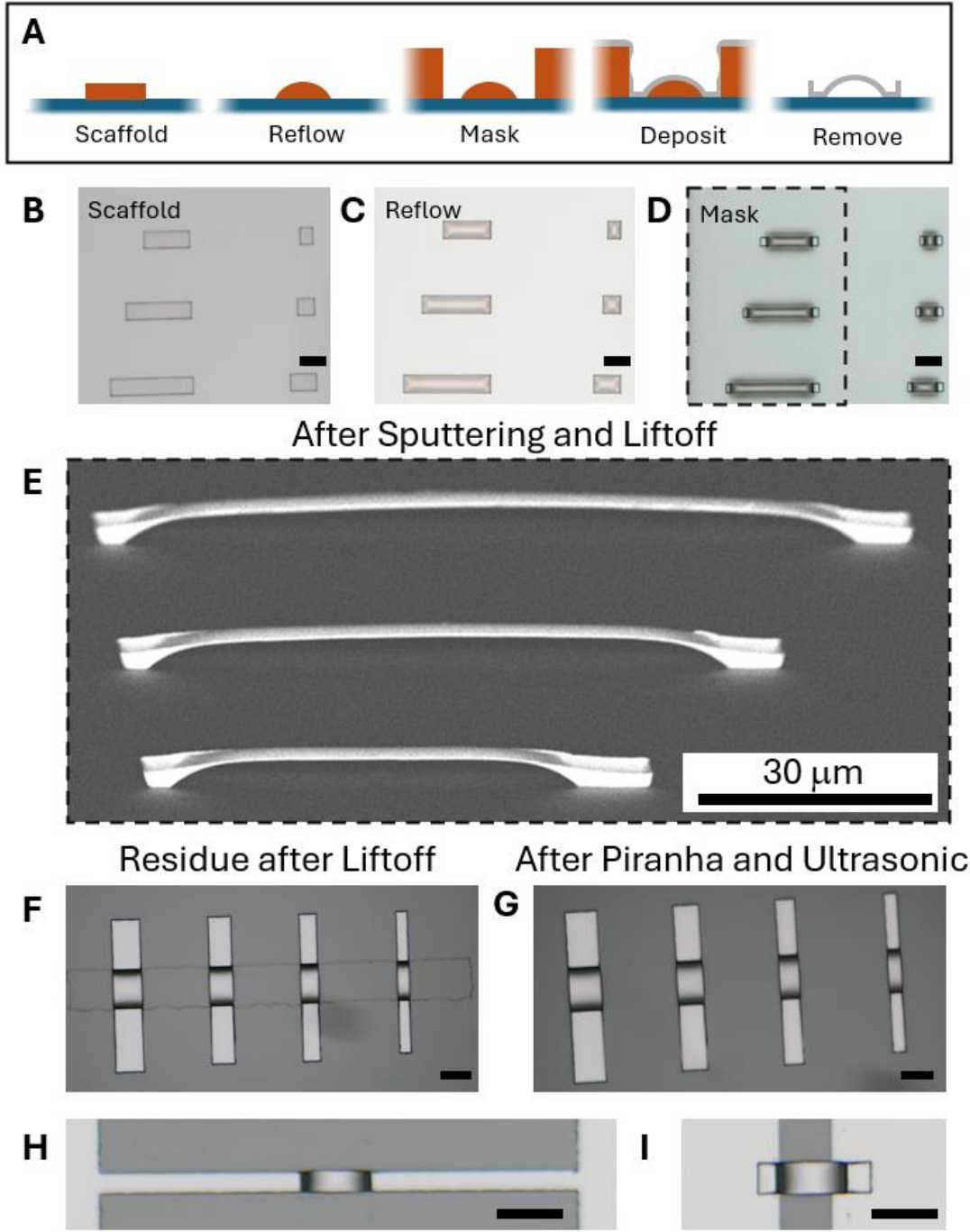

**Figure 2. Growth of α-Ta on photoresist enables airbridge definition.** (A) Schematic of Ta airbridge fabrication process: A photoresist scaffold is patterned, then heated (reflowed) to round it. Next, a second lithography step defines a liftoff mask, the Nb/Ta deposit is performed, and finally the resist removed by solvents and Piranha solution. (B) Microscope image of the scaffold before and (C) after reflow. (D) Image after the mask is developed. (E) A tilted SEM image of bridge structures after the resist removal step. (F-G) Piranha etching completely removes a stubborn residue left behind by the scaffold (H-I) Devices for transport measurement. (All images are optical apart from E. All scale bars are 30 μm)

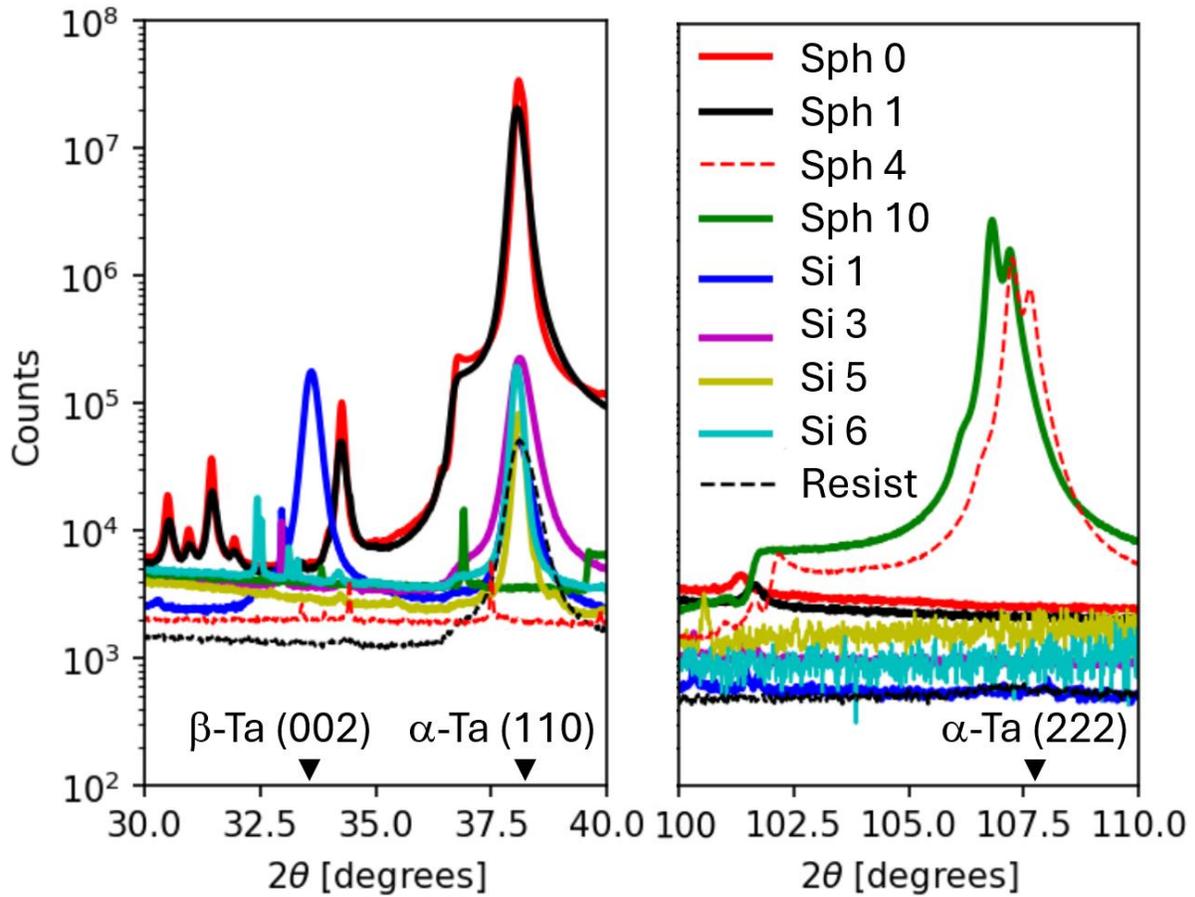

**Figure 3. X-ray diffraction (θ-2θ with continuous ϕ rotation) of Ta films grown on sapphire, silicon, and AZ1518 photoresist.** [111]-oriented Ta (Sph 10, Sph 4) shows the sharpest reflections and almost no features in the region characteristic of [110]-oriented films (Sph 0,1 and Si 3-6), and conversely none of the other films show a (111) reflection.

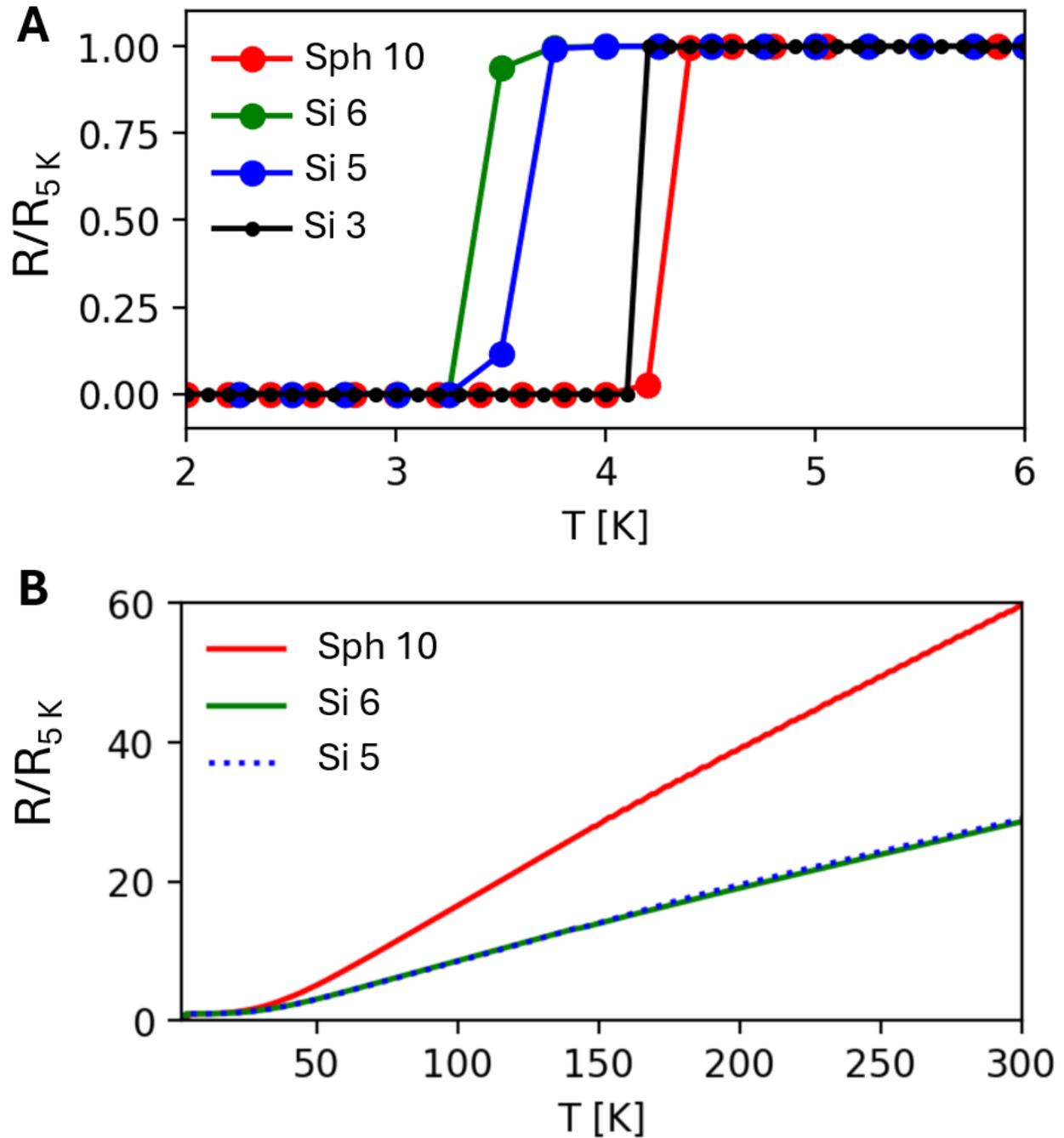

**Figure 4. Variable-temperature transport measurements.** DC resistance measurements show (A) the highest critical temperatures to be reached by Ta films grown at high temperature on sapphire (Sph 10), room-temperature-nucleated Ta to have a slightly diminished $T_c$ (Si 3) and films grown under the same high temperature conditions but on Si to have a lower $T_c$ and (B) resistivity ratio (Si 5, 6).

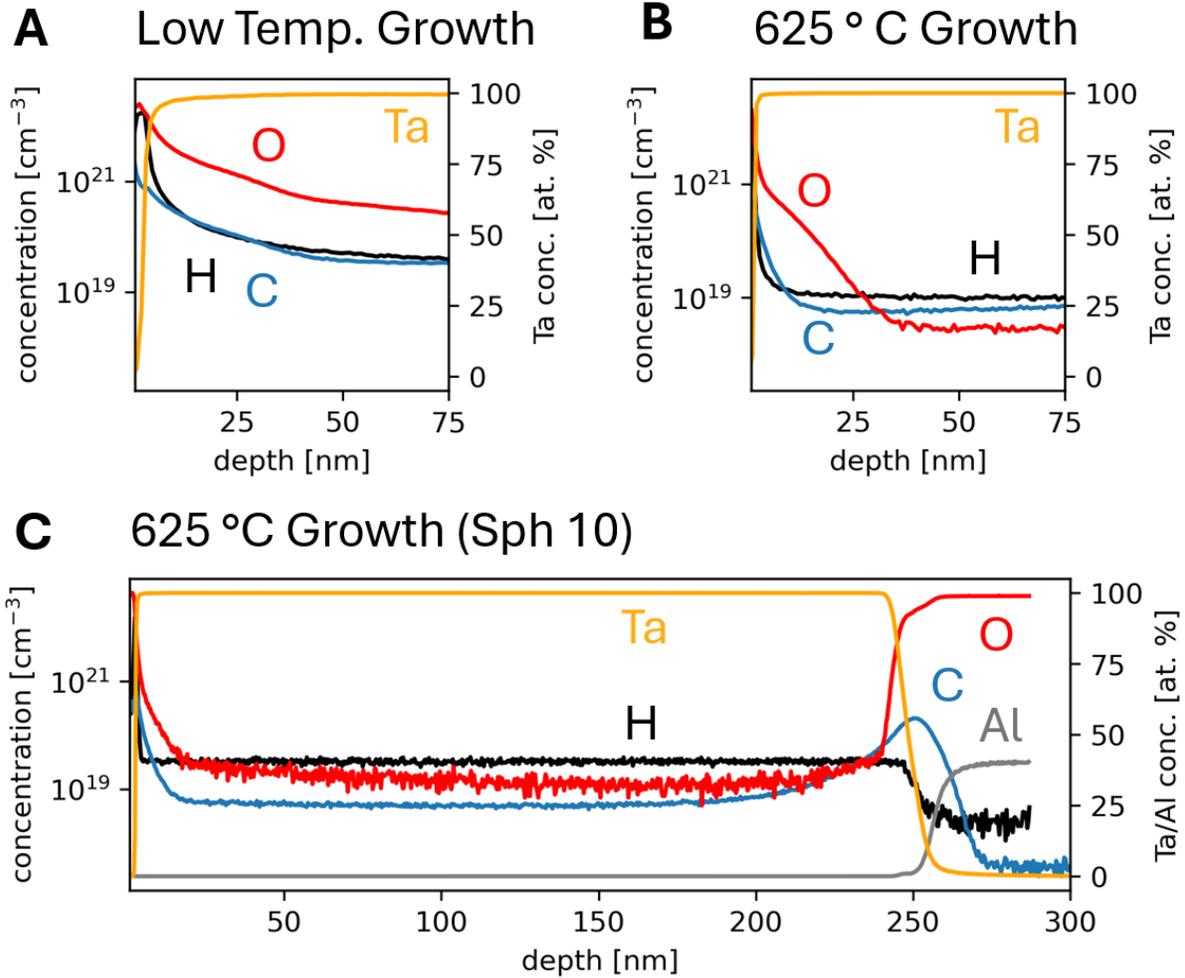

**Figure 5. Secondary ion mass spectroscopy (SIMS) of films grown at high and low temperatures.** (A) SIMS of the surface of a film nucleated at room temperature showing H, C and O contaminants to extend further into the film surface as compared to a film (B) grown at high temperature (this film was grown under conditions similar to films Sph 6-12). (C) In Sph 10, which showed a high residual resistivity ratio and resonator internal quality factors, there remain substantial quantities of H, C, and O as compared to the sapphire substrate.

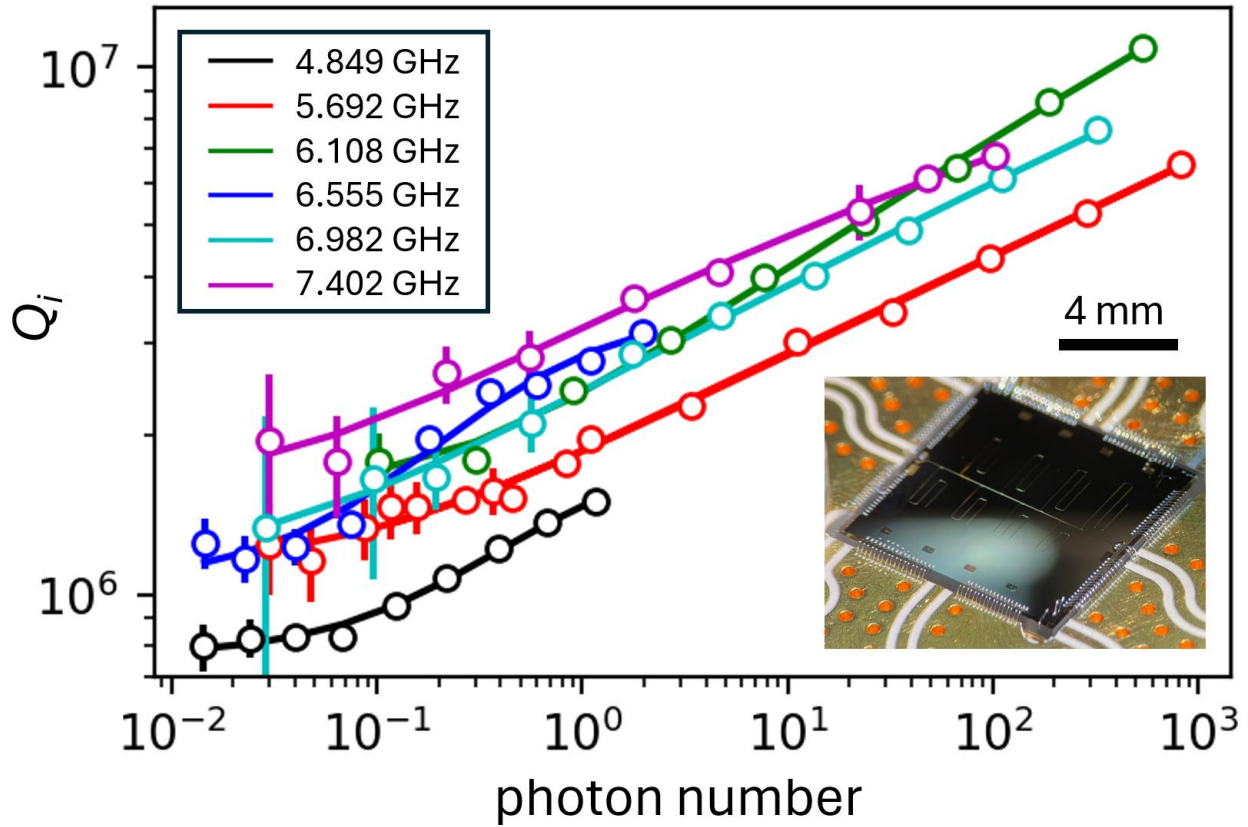

**Figure 6. Resonator Measurements.** Internal quality factor $Q_i$ of coplanar waveguide resonators patterned from Sample Sph 10, measured at 10 mK, labeled by the resonator center frequency. The center conductor of the resonators is 6 μm wide and the gap is 3 μm.[44] The error bars represent the uncertainty in the internal quality factor fit parameter. Inset, photograph of the resonator die.

**Supplementary Material**

Additional XRD and microscopy data are included in the supplementary material.

**Acknowledgements**

This material is based upon work supported by LLNL-LDRD-24-ER-045, (supporting contributions by J.L.D. and E. K.); by AFOSR MQC (supporting the contributions of L.D.A. and A.A.); and by U.S. Department of Energy, Office of Science, Basic Energy Sciences, under award number DE-SC0020313 (supporting the contributions of S.I., P.M.V.). S.I. and P.M.V. acknowledge the use of facilities supported by the Wisconsin MRSEC (DMR-2309000). This work was performed in part under the auspices of the U.S. Department of Energy by Lawrence Livermore National Laboratory under contract DE-AC52-07NA27344. Lawrence Livermore National Security, LLC.

**Author Declarations**

*Conflict of Interest*

LDA has patent pending to Lawrence Livermore National Security, LLC. All other authors declare that they have no known competing financial interests or personal relationships that could have appeared to influence the work reported in this paper.

*Author Contributions*

**L.D.A.:** Conceptualization, Investigation, Data curation, Formal analysis, Writing - original draft, Supervision. **E.K.:** Investigation (XRD,PPMS). **S.I.:** Investigation (STEM). **A.A.:** Investigation (XRD, PVD), Methodology, Writing - review & editing. **P.M.V.:** Supervision, Writing - review & editing. **V.L.:** Funding acquisition, Writing – Review & Editing, Conceptualization. **J.L.D:** Conceptualization, Supervision. **Y.J.R.:** Formal analysis.

**Data Statement**

The data that support the findings of this study are available from the corresponding author upon reasonable request.

# Supplementary Material: Growth and Structure of alpha-Ta films for Quantum Circuit Integration

Loren D. Alegria[1]*, Alex Abelson[1], Eunjeong Kim[1], Soohyun Im[2], Paul M. Voyles[3], Vincenzo Lordi[1], Jonathan L Dubois[1], Yaniv J. Rosen[1]

[1] Physics Division, Lawrence Livermore National Laboratory, Livermore, CA, 94550, USA

[2] Department of Materials Design and Innovation, University at Buffalo, Buffalo, NY, 14260, USA

[3] Department of Materials Science and Engineering, University of Wisconsin-Madison, Madison, WI, 53706, USA

*corresponding author: alegria4@llnl.gov

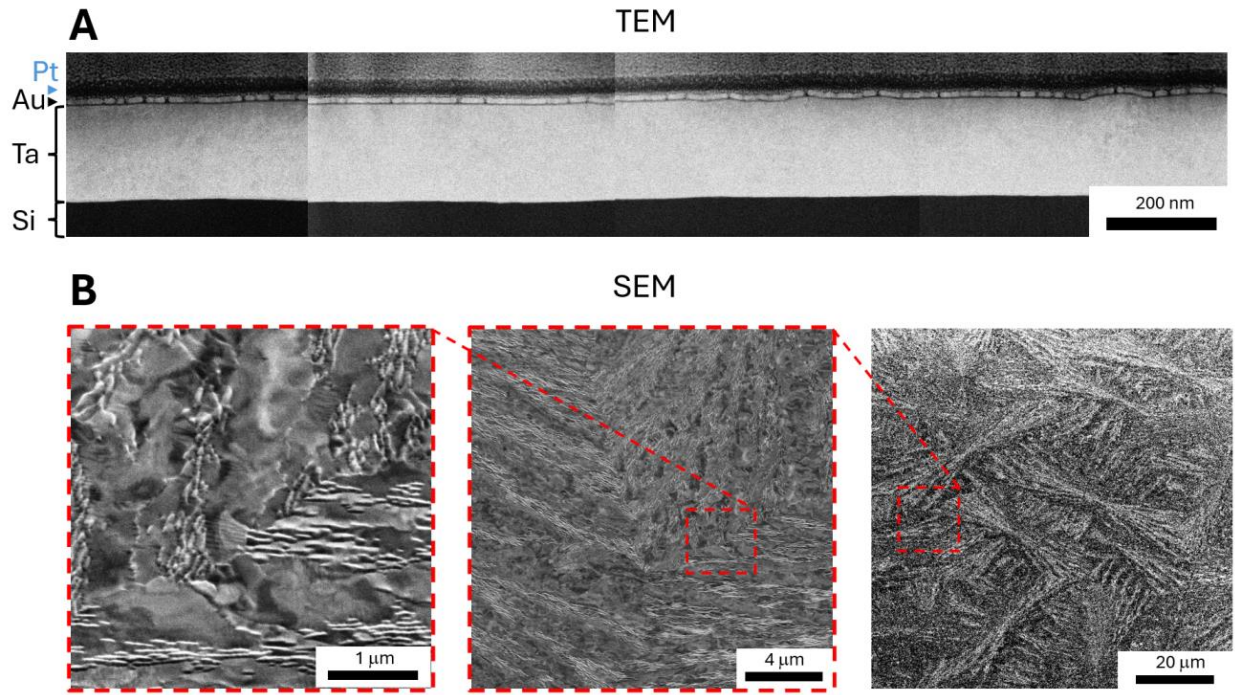

**Figure S1. Grain texturing of Ta on Si (Sample Si 6).** (A) Relatively smooth films grow on Si under the conditions described in the main text, as seen in the wide-field STEM image here. (Note that the layers on the top of the Ta film are Au and Pt deposited during the TEM sample preparation.) (B) Nonetheless the films Si 5 and Si 6 displayed distinctive texturing spanning a ~ 100 um length scale. As can be seen in the SEM images, the dendritic structure consists of alternating bands of wide and narrow grains. The largest structures, as in the rightmost image, were very uniformly distributed over the wafer above this scale. The structure may originate in electric fields present in the films during growth.

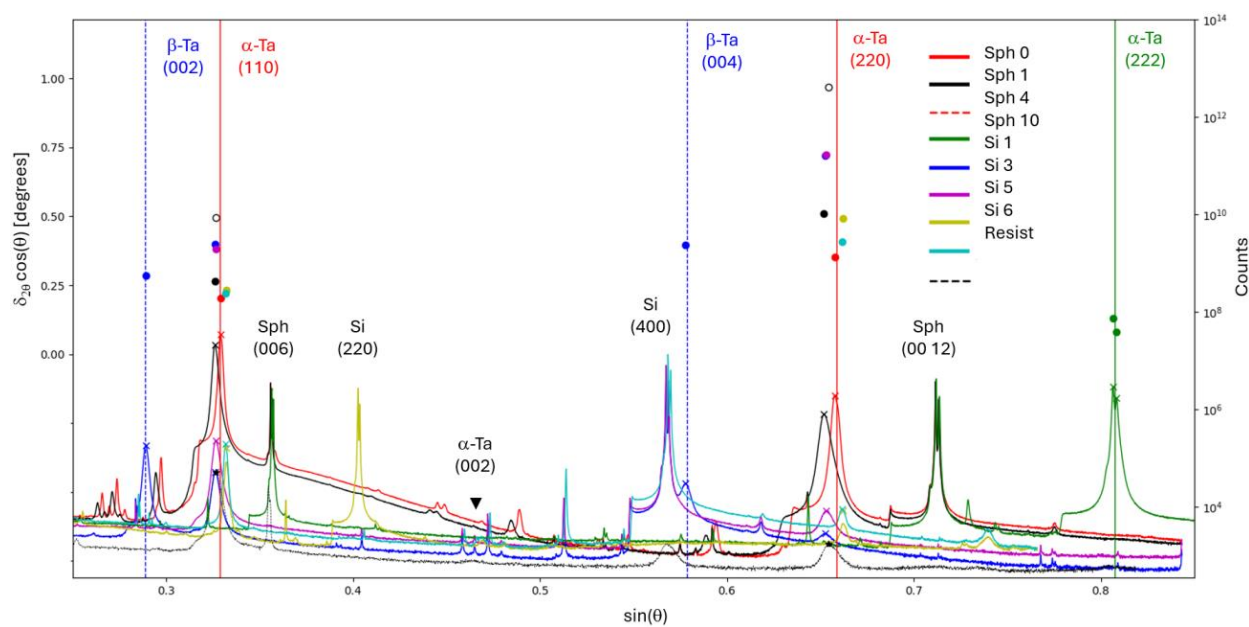

**Figure S2. Williamson-Hall plot and complete XRD data.** The full XRD spectra are plotted as a function of $sin(\theta)$, emphasizing the relative contributions of strain and grain size broadening of the $\alpha$-Ta reflections. The circles represent the FWHM of the peaks, scaled by the cosine [$\delta_{2\theta} cos(\theta)$], and are plotted on the left hand axis. The (220) reflections are considerably broadened at the higher angle. By contrast, the (222) reflections of Sph 10 and Sph 4 remain sharp at the high angle where they occur.

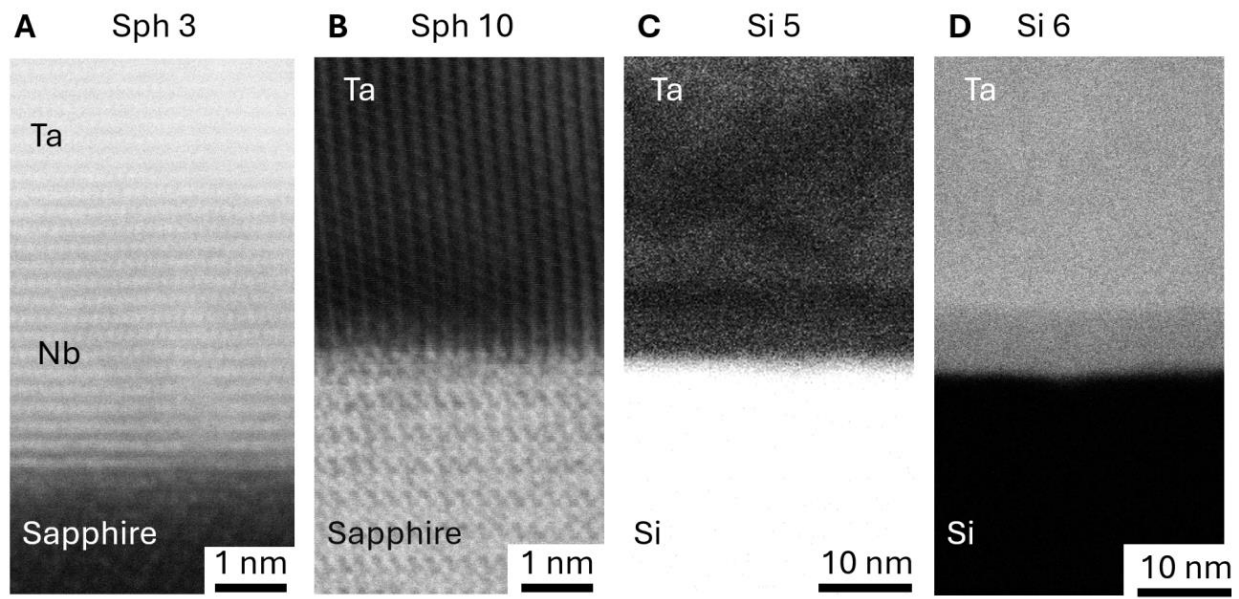

**Figure S3. Cross-sectional STEM Studies of α-Ta films on Si and Sapphire.** (A) Films nucleated by Nb at room-temperature as described in the text show sharp interface to the sapphire, a clear Nb boundary layer and continuous epitaxy through the Nb-Ta interface. (B) In films grown at high temperature, there remains an atomically sharp interface to the sapphire. On (100) and (110) Si (C and D, respectively) a boundary layer of 8-10 nm has a Si-Ta composition as determined by EDS.

| Resonator Number | $f_0$ (GHz) | $Q_{TLS}^{-1} \times 10^6$ | $n_C$ | $\beta$ | $Q_h^{-1} \times 10^6$ |
|---|---|---|---|---|---|
| 1 | 4.849 | 1.22 | 0.087 | 0.6 | 0.1 |
| 2 | 5.692 | 0.87 | 0.082 | 0.37 | 0.00 |
| 3 | 6.108 | 0.62 | 0.24 | 0.54 | 0.01 |
| 4 | 6.555 | 1.0 | 0.016 | 0.54 | 0.03 |
| 5 | 6.982 | 0.83 | 0.03 | 0.43 | 0.02 |
| 6 | 7.402 | 0.6 | 0.036 | 0.47 | 0.06 |

**Table ST1: Resonator Quality Factor Fit Results.** For each resonator of Fig. 6 of the main text, $Q_i(n)$ data are fit according to the form $Q_i^{-1} = Q_{TLS}^{-1}(1+n/n_C)^{-\beta/2} + Q_h^{-1}$ where $Q_{TLS}$ is the TLS-limited internal quality factor, $n_C$ is the critical photon number, the exponent $b$ is device dependent and typically taken as a free parameter, and $Q_h$ is the quality factor in the high power limit.

### *Methods: Resistance vs Temperature Measurements*

The four-terminal DC resistance measurements of Figure 4 in the main manuscript were performed in a Quantum Design PPMS cryostat, linearly ramping the temperature from 2 K to 300 K.